\documentclass[12pt,thmsa]{article}
\usepackage{amssymb}
\usepackage{graphicx}

\begin{document}

\author{Alfredo B. Henriques\thanks{%
email:alfredo@fisica.ist.utl.pt} \\
CENTRA/Dep. de F\'{i}sica\\
Instituto Superior T\'{e}cnico\\
Av. Rovisco Pais, 1049-001 Lisboa, Portugal}
\title{The Stochastic Gravitational-Wave Background and the Inflation to Radiation
Transition in the Early Universe}
\date{June 2007 }
\maketitle

\begin{abstract}
In this paper we calculate the gravitational-wave power-spectra
corresponding to different types of transitions between the inflationary
regime and the radiation era. We study four cases, where inflation is
followed by a stiff matter phase, or by a dust dominated phase, or by a
combination of the two, before the universe enters the radiation era. Use is
made of differential equations for the Bogoliubov coefficients, allowing us
to model all the transitions between the different eras as continuous. New
features appear, for frequencies above 10$^{7}$ rad s$^{-1}$ in an otherwise flat
part of the spectrum.
\end{abstract}

I. \underline{Introduction}

Although gravitational-waves of cosmological origin have not yet been
directly detected, they are at present the object of an important effort of
research, as they provide us with a unique telescope to the very early
stages of the formation of the universe. Using the upper bound set up by the
COBE measurement of the total flux of energy coming from the metric
perturbations \cite{b10}, we explore the consequences, to the power spectrum
of the stochastic background of gravitational-waves, of the possible
existence of different regimes of transition between the inflationary
universe and the standard radiation dominated expansion. This transition is
far from being well known and it may be of some interest to explore the
signatures associated with different possibilities, just assuming that an
ideal detector will be able, in the more or less distant future, to explore
in some detail the power spectrum. Examples of this can be seen in the works
of Giovannini (\cite{b8}, \cite{b9} and \cite{b11}), as well as in Mendes
and Liddle concerning models with thermal inflation \cite{b7}.

In the present work, and contrary to these authors, we do not use the sudden
transition approximation between different eras. We model the transitions as
continuous and integrate numerically the differential equations, derived by
Parker \cite{b1}, for the time dependent Bogoliubov coefficients. Associated
with the sudden transition approximation we always have an overproduction of
gravitons of large frequencies, requiring the introduction of an explicit
cut-off for frequencies larger than the rate of expansion at the time of the
transition. Quite often, also, some more or less delicate cancellations have
to be worked out in the limit of the low frequencies. All this is
automatically taken into account by the use of continuous transitions.

For simplicity, we assume that inflation is dominated by a cosmological
constant, after which we have a transition period modelled by a
self-interacting scalar field, whose energy is continuously transferred to a
fluid, with an equation of state of the form $p=\gamma \varrho $, through
the action of a frictional term. This is followed by the radiation and
matter eras. At a certain point, during the matter era, the universe is
supposed to become dominated by dark energy, described by an equation of
state of the form $p_{de}=w\varrho _{de}$, with $w$ negative,in agreement with the results of
the WMAP\ satellite\cite{b2}.

In the next section we introduce our model and derive the main equations. In
section III we describe the results of the numerical simulations for the
four cases investigated and the corresponding spectra. In the first example,
after inflation we have a period dominated by a stiff matter fluid; in the
second example inflation is followed by a dust dominated period. In the
other two examples, inflation is followed by a combination of stiff matter
and dust. All the cases end with the universe entering the standard
radiation era. In section IV we make some additional comments and discuss
the possibility of detecting the effects described.

\bigskip

II. \underline{The model}

The background metric is defined by

\begin{equation}
ds^{2}=-a^{2}(\eta )(-d\eta ^{2}+\delta _{ij}dx^{i}dx^{j})  \label{eq1}
\end{equation}

At the beginning the expansion is driven by a cosmological constant, up to a
time $\eta =\eta _{i}$, when both the transition period and our integration
begin. The scale factor up to $\eta =\eta _{i}$ is given by

\begin{equation}
a(\eta )=\frac{1}{H(\eta _{1}-\eta )}\rm{ , }\eta <\eta _{i}  \label{eq2}
\end{equation}
where $H$ is the Hubble constant during inflation and $\eta _{1}=a(\eta
_{i})/a^{\prime }(\eta _{i})+\eta _{i}$ \cite{b3}. In conformal time the
Einstein's equation for the scale factor is

\begin{equation}
a^{\prime \prime }=\frac{4\pi G}{3}a^{3}(\varrho -3p)  \label{eq3}
\end{equation}
where

\begin{equation}
\varrho _{total}=\frac{1}{a^{2}}(\frac{1}{2}\varphi ^{\prime
2}+a^{2}V(\varphi ))+\varrho  \label{eq4}
\end{equation}
and

\begin{equation}
p_{total}=\frac{1}{a^{2}}(\frac{1}{2}\varphi ^{\prime 2}-a^{2}V(\varphi ))+p
\label{eq5}
\end{equation}
define the total energy density and pressure, and $\varrho $ and $p=\gamma
\varrho $ are the energy density and pressure of the fluid into which the
scalar field energy density is transferred. For numerical convenience, we
divide the integration into two parts, the first part taking us from $\eta
_{i}$ up to $\eta _{1}$, the end of the transition period, also marking the
beginning of radiation domination. At this point, the energy that initially
was in the scalar field will have been transferred to the radiation field $%
p=\varrho /3$ (the energy contained in the matter and dark energy components
are still completely negligible). Once the present content of the universe
is known from the WMAP measurements, we know the redshift corresponding to
this point and, from it, we easily find the redshift corresponding to $\eta
=\eta _{i}$. We then begin the second part of our integration that will take
us from $\eta _{1}$ till the present time, $\eta _{o}$, using the Einstein's
equation in the form

\begin{equation}
a^{\prime \prime }=\frac{4\pi G}{3}a^{3}(\frac{\varrho _{mo}}{a^{3}}+(1+3|w|)%
\frac{\varrho _{deo}}{a^{3(1-|w|)}})  \label{eq6}
\end{equation}
where, in conformal coordinates, the radiation fluid does not contribute to
it and $\rho _{mo}$ and $\varrho _{deo}$ are the present values of the
energy densities in the matter and dark energy components. We define the
scale factor at the present time to be $a(\eta _{o})=1$; knowing $\varrho
_{mo}$ and $\varrho _{deo}$, we know the redshifts corresponding to the
transitions between radiation to matter domination and from matter to dark
energy.

The other background equations, for the first stage of integration, are the
equation for the scalar field

\begin{equation}
\varphi ^{\prime \prime }+2\frac{a^{\prime }}{a}\varphi ^{\prime
}+a^{2}\partial _{\varphi }V(\varphi )=-a\Gamma _{\varphi }\varphi ^{\prime }
\label{eq7}
\end{equation}
and the equation(s) for the fluid(s) $p=\gamma \varrho $. If, after the end
of inflation at $\eta =\eta _{i}$, and before the radiation stage ($\gamma
=1/3$), we have an intermediate stage with $p=\gamma \varrho $ ($\gamma \neq
1/3$), we then have the two equations

\begin{equation}
\varrho ^{\prime }=-3\frac{a^{\prime }}{a}(1+\gamma )\varrho +\frac{1}{a}%
\Gamma _{\varphi }\varphi ^{\prime 2}-l.a.\Gamma \varrho  \label{eq8}
\end{equation}
and

\begin{equation}
\varrho _{r}^{^{\prime }}=-4\frac{a^{\prime }}{a}\varrho _{r}+a\Gamma \varrho
\label{eq9}
\end{equation}
$l$ being a control number, which is different from zero only after a
pre-defined and more or less extended interval of time such that, by then,
the energy of the scalar field has effectively been transferred to the fluid 
$p=\gamma \varrho $; $\Gamma _{\varphi }$ and $\Gamma $ are the frictional
decay constants that control this tranfer of energy from the scalar field to
the fluid and from this to radiation field, respectively. The remaining
equations to be used in both stages of integration are the ones for the
Bogoliubov coefficients, defined below (equations \ref{eq18} and \ref{eq19}). We
must stress that our aim is not to construct a truly realistic model for the
transition period marking the end of inflation, but to explore different
possibilities with the help of a simple and reasonable toy-model.

As for the tensor perturbations $h_{ij}$ (latin indices from 1 to 3) of the
space part of the metric $a(\eta )^{2}(\delta _{ij}+h_{ij}(\eta ,\mathbf{x}%
)) $, they can be expanded, in the usual manner, in terms of plane-waves

\begin{eqnarray}
h_{ij}(\eta ,\mathbf{x)} &=&\rm{{}}\sqrt{8\pi G}%
\sum_{p=1}^{2}\int \frac{d^{3}k}{(2\pi )^{3/2}a(\eta )\sqrt{2k}}\nonumber \\
&&\times [a_{p}(\mathbf{k},\eta )\varepsilon _{ij}(\mathbf{k},p)e^{i\mathbf{k%
}.\mathbf{x}}\xi (\mathbf{k},\eta )+herm.conj.]  \label{eq10}
\end{eqnarray}
where \textbf{x} denotes the spatial coordinates, $k=\mid \mathbf{k\mid =}%
2\pi a/\lambda =\omega a$; the index $p$ runs over the two polarizations of
the gravitational-waves, and $\varepsilon _{ij}$ is the polarization tensor, 
$a_{p}$ the annihilation operator and $\xi $ the mode function for the
gravitational-waves. The mode function obeys the equation

\begin{equation}
\xi ^{\prime \prime }+(k^{2}-a^{\prime \prime }/a)\xi =0,  \label{eq11}
\end{equation}
the derivatives being with respect to conformal time $\eta $. We now express
the creation and annihilation operators, at $\eta ,$ in terms of the initial
creation and annihilation operators $A_{p}^{\dagger }(\mathbf{k})$ and $%
A_{p}(\mathbf{k})$, through the Bogoliubov coefficients $\alpha (k,\eta )$
and $\beta (k,\eta )$:

\begin{equation}
a(\mathbf{k},\eta )=\alpha (k,\eta )A(\mathbf{k})+\beta ^{*}(k,\eta
)A^{\dagger }(\mathbf{k})  \label{eq12}
\end{equation}
$\alpha $ and $\beta $ satisfying the relation

\begin{equation}
|\alpha |^{2}-|\beta |^{2}=1\rm{.}  \label{eq13}
\end{equation}

These coefficients, which, in the sudden approximation, are calculated by
requiring the mode functions and their derivatives to be continuous across
the transitions, obey the set of coupled of differential equations \cite
{b1,b3}

\begin{equation}
\alpha ^{\prime }(\eta )=\frac{i}{2k}(\alpha (\eta )+\beta (\eta
)e^{2ik(\eta -\eta _{o})})\frac{a^{\prime \prime }}{a}  \label{eq14}
\end{equation}
and

\begin{equation}
\beta ^{\prime }(\eta )=-\frac{i}{2k}(\beta (\eta )+\alpha (\eta
)e^{-2ik(\eta -\eta _{o})})\frac{a^{\prime \prime }}{a}  \label{eq15}
\end{equation}
where $\eta _{o}$ is an arbitrary constant to be put equal to $\eta _{i}$;
in Parker's integral equations the function $W(k,\eta )$ has here been
replaced by the ansatz $W(k,\eta )=k$. Introducing the functions $X(k,\eta )$
and $Y(k,\eta )$ through the definitions

\begin{equation}
\alpha =\frac{1}{2}(X+Y)e^{ik(\eta -\eta _{o})}  \label{eq16}
\end{equation}

\begin{equation}
\beta =\frac{1}{2}(X-Y)e^{-ik(\eta -\eta _{o})}  \label{eq17}
\end{equation}
the two equations above take the form

\begin{equation}
X^{\prime \prime }+(k^{2}-\frac{a^{\prime \prime }}{a})X=0  \label{eq18}
\end{equation}

\begin{equation}
Y=\frac{i}{k}X^{\prime } \label{eq19}
\end{equation}
these being the remaining equations of our system. When the scale factor
obeys a simple power-law, the solution to equation (\ref{eq18}) is expressed in
terms of Hankel functions. When this is not so, as in the present situation,
we have to integrate it numerically, with appropriate initial conditions.
For the case of an initial stage dominated by a cosmological constant and a
scale factor $a$ defined by equation (\ref{eq2}), the solution is well-known and
will define the initial conditions for $X$ and $X^{\prime }$:

\begin{equation}
X=1-\frac{i}{k(\eta _{1}-\eta _{i})}  \label{eq20}
\end{equation}

\begin{equation}
\frac{i}{k}X^{\prime }=Y=1-\frac{1}{k^{2}(\eta _{1}-\eta _{i})^{2}}-\frac{i}{%
k(\eta _{1}-\eta _{i})}  \label{eq21}
\end{equation}
with $(\eta _{1}-\eta _{i})$ already known. At the end of the integration,
when the initial and final states are characterized by different kinds of
mode functions, an additional Bogoliubov transformation is necessary, as
explained in references \cite{b3} and \cite{b4}; with a final stage
dominated by dark energy with an equation of state $p_{de}=w\varrho _{de}$, 
$w\approx -0.78$, we may ignore the final projection, as it is numerically
irrelevant.

Once we have $X$ and $X^{\prime }$ at $\eta =\eta _{o}$, $\alpha _{final}$
and $\beta _{final}$ can be recovered from equations (\ref{eq16}), (\ref{eq17})
and (\ref{eq19}). The number of gravitons created during the expansion of
the universe is given by $|\beta _{final}|^{2}$ \cite{b1}; taking into
account that the density of states is $\omega ^{2}d\omega /(2\pi ^{2}c^{3})$
and that each graviton contributes with two polarization states $2\hbar
\omega $, then, from the definition of the energy density $dE=P(\omega
)d\omega $, we have the following expression \cite{b5} for the
power-spectrum $P(\omega )$:

\begin{equation}
P(\omega )=\frac{\hbar \omega ^{3}}{\pi ^{2}c^{3}}|\beta
_{final}|^{2}  \label{eq22}
\end{equation}
in units erg s cm$^{-3}$. We can also express the results in terms of the relative logarithmic
energy-spectrum of the gravitational-waves, which is defined by

\begin{equation}
\Omega (\omega ,\eta _{p})=\frac{1}{\rho _{c}}\frac{d\rho _{gw}}{d\ln \omega 
}  \label{eq23}
\end{equation}
where $\rho _{c}$ is the value of the present time critical density and $%
\rho _{gw}$ is the gravitational-wave energy density

\begin{equation}
\rho _{gw}=\int P(\omega )d\omega .  \label{eq24}
\end{equation}
When performing these calculations within the sudden transition
approximation, due to a logarithmic divergence, a cut-off needs to be
introduced in equation (\ref{eq24}) on the high-frequency gravitons, which
depends on the speed of the physical transition \cite{b5}. This is not
necessary when the transitions are continuous, as shown in \cite{b3} and 
\cite{b6}. In the low-frequency limit we have a natural cut-off provided by
the size of the present horizon \cite{b5}.

The gravitational-wave contribution cannot be larger than the total amount
of anisotropy measured by COBE. This implies that $\Omega $ is limited by
(value taken from \cite{b8} )

\begin{equation}
\Omega \lesssim 1.37\times 10^{-10}  \label{eq25}
\end{equation}
for frequencies corresponding to the present horizon size $\omega _{hor}$,
where we used $H=71$ km s$^{-1}$ Mpc$^{-1}$, giving $\omega _{hor}=
10^{-17}$ rad s$^{-1}$. In the numerical simulations, we used an inflationary scale
chosen in order to avoid any conflict with the known limits for the scalar
perturbations.

\bigskip

III. \underline{Numerical simulations}

To proceed with the numerical calculations, we used the simple scalar field
potential

\begin{equation}
V(\varphi )=V_{o}(\frac{\varphi }{\mu })^{p}  \label{eq26}
\end{equation}
taking $p=4$ and $\mu =\varphi (\eta _{i})$, the initial value of $\varphi $%
. Thus, at the beginning of the integration $V=V_{o}$, $V_{o}$ being then
equal to the scale of the inflationary regime. We now give and discuss a few
examples.

\bigskip

\underline{Example 1}

In the first example we assume that the inflationary regime is followed
by a period dominated by a fluid with a stiff matter equation of state, equation (%
\ref{eq9}) with $\gamma =1.$

An instance of a model of this kind is the quintessential inflationary model
developed by Peebles and Vilenkin \cite{b12} and explored, from the point of
view of the production of gravitational waves, by Giovannini \cite{b9}.
First, the energy in the scalar field is converted into a stiff matter
fluid, by the action of the $\Gamma _{\varphi }$ term; we consider the
conversion concluded when $\varrho _{stiff}a^{6}$ becomes constant. Then,
after a period of time, which we took, in the present instance,
approximately equal to $\Delta \eta =4.5\times 10^{5}$, in Planck units, the
stiff matter is in turn converted into radiation with the help of the
phenomenological frictional term $\Gamma _{stiff}$ in equations (\ref{eq8}) and (%
\ref{eq9}). We used as initial conditions, at $\eta =\eta _{i},$ the values 
$\varphi _{i}=0.001$, $\varphi _{i}^{\prime }=-0.00001$, $\Gamma _{\varphi
}=\Gamma _{stiff}=0.01$, all in Planck units; we also fixed the scale of
inflation $(V_{o})^{1/4}=4.3\times 10^{16}$ GeV. Finally, from the results
of the WMAP we fixed the equation of state of the dark energy component as $%
p=-0.78\varrho $ and $\Omega _{de}=0.73$.

With these values, the redshift corresponding to $\eta _{i}$ is $z_{\inf
}=8.53\times 10^{28}$. The relative logarithmic energy-spectrum $\Omega
(\omega )$ is given in figure 1. We see the new feature, first noticed by
Giovannini \cite{b9}, appearing in the hard branch of $\Omega (\omega ),$
for frequencies $\geq 10^{7}$ rad s$^{-1}$. From the results in \cite{b3}, we
expect a cut-off to emerge naturally for frequencies $\omega \succeq
10^{12}$ rad s$^{-1}$, sensitive to the first transition between the end of inflation
and the stiff matter stage.

The jump in $\Omega (\omega )$ not only is sensitive to the scale $%
(V_{o})^{1/4}$, as it also increases with the increase in the interval of
time $\Delta \eta $, during which we allow the stiff matter stage to
dominate. In the present case we have, for instance, $\Omega (\omega
_{hor})=1.3\times 10^{-11}$, $\Omega (\omega =10^{-12}\mbox{rad s$^{-1}$})=2.7\times
10^{-13}$, in the horizontal part of the spectrum, and $\Omega (\omega
=10^{11}\mbox{rad s$^{-1}$})=9.6\times 10^{-13}$, at the peak.

For comparison we added the curve corresponding to the sudden transition
approximation, using equations (4.8) in \cite{b5}. Being a different model,
we had to fix the redshift for the transition from the radiation to the
matter era. We used the value $z=2.2\times 10^{3}$, midpoint between the
values given in \cite{b2} for $z_{eq}$ and $z_{dec}$. In each of the examples,
figures 1-4, the Hubble constant at the sudden transition between inflation 
and radiation was chosen in order for $\Omega (\omega _{hor})$ 
to  have  the same value as for the corresponding continuous transition case. 
This required a small adjustment in its value, explaining the slight 
difference in the horizontal line in the different plots.

\bigskip

\underline{Example 2}

This time, following the inflationary period, we have a period with a
dust-like equation of state $p=0$.

We keep the initial conditions of the preceeding example, except for $\Gamma
_{d}=0.01$ and $(V_{o})^{1/4}=$ $6.2\times 10^{16}$ GeV, giving $\Omega
(\omega _{hor})=5.6\times 10^{-13}$. The extension of the matter dominated
period (during which $\varrho _{d}a^{3}$= constant) is $\Delta \eta
=1.4\times 10^{5}$. We have $z_{\inf }=8.\times 10^{29}$.

Instead of an increase in $\Omega $, we find a decrease (see figure 2),
which will also depend on $\Delta \eta $. In the present case $\Omega
(\omega =10^{10}\mbox{rad s$^{-1}$})=6.14\times 10^{-16}$, lower than the
value $\Omega (\omega =10^{-12}\mbox{rad s$^{-1}$})=6.7\times 10^{-15}$. Although during
the dust matter interval gravitons are produced, particularly during the
transitions to and from this period, the increase in the scale factor $a$ is
faster than in the previous example, leading to a stronger redshift of the
energy, and a decrease in $\Omega $. The increase around $\omega=10^{12}$ rad s$^{-1}$ is due to the presence of the $V(\varphi)$ potential.

In the same spirit of exploring different configurations between inflation
and radiation, we next investigate what happens when we have a period with
stiff matter followed by dust, or vice-versa, before the final transition
into the radiation era.

\bigskip

\underline{Example 3}

Inflation is now followed by a dust-like period $(\gamma =0)$, which is in
turn followed by stiff matter $(\gamma =1)$, before the final transition
into the radiation era.

The dust and stiff matter periods have an extension given by $\Delta \eta
_{dust}=1.8\times 10^{5}$ and $\Delta \eta _{stiff}=1.4\times 10^{5}$,
respectively. The energy in $\varphi $ is converted into dust with the help
of $\Gamma _{\varphi }$, then dust is converted into stiff matter by the
action of $\Gamma _{d}$ and, finally, stiff matter into radiation via $%
\Gamma _{stiff}$; we keep $\Gamma _{\varphi }=\Gamma _{d}=\Gamma
_{stiff}=0.01$. The scale of inflation $(V_{o})^{1/4}=7.6\times 10^{16}$ GeV
and $\Omega (\omega _{hor})=7.2\times 10^{-13}$; z$_{\inf }=7.5\times
10^{29}$.

The results are shown in figure 3. We notice that, in the range of
frequencies $\omega \geq 10^{7}$ rad s$^{-1}$, we first have an increase, due to
the extra power brought in by the stiff matter component, and then the dip
caused by the dust stage. It is illustrative to compare these results with
those from example 4.

\bigskip

\underline{Example 4}

We exchange the two intermediate periods, first a stiff matter period and
then dust.

Taking $\Delta \eta _{stiff}=6.\times 10^{5}$, $\Delta \eta _{d}=1.4\times
10^{6}$ and $(V_{o})^{1/4}=4.8\times 10^{16}$ GeV, with the same $\Gamma
^{\prime }$s, we find the results seen in figure 4, first a dip and then an
increase with respect to the horizontal line, this last at the level of $%
\Omega (\omega =10^{-12}\mbox{rad s$^{-1}$})=4.1\times 10^{-13}$.

In the examples above we used values for $(V_{o})^{1/4}$ between $5$ and $%
7.6\times 10^{16}$ GeV, in reasonable agreement with the approximate limit $%
2.6\times 10^{16}$ GeV found in \cite{b13} . Given the uncertainties around
this period, we believe these values for $(V_{o})^{1/4}$ to be quite
reasonable.

\bigskip

Other, more complex, situations could have been devised, but these examples
already prove that, if we have a complicated transition between inflation
and radiation, and an \textit{ideal} detector, by observing the shape of $\Omega
(\omega )$ we may eventually be able to say a lot about such transitions, as
the changes in $\Omega (\omega )$ not only reflect the different equations
of state, but also the order and sequence of the different periods. This can
be seen from the comparison between figures 3 and 4. In all cases studied,
the size of the new features depends on the extension of the different
intermediate periods and the scale taken for the inflationary period.

\medskip

\medskip

IV. \underline{Concluding remarks }

In this paper we developed a simple toy model to study different types of
transitions between the inflationary regime and the radiation era and, with
the help of differential equations for the Bogoliubov coefficients,
calculated the corresponding gravitational-wave power-spectra. The usual
sudden transition between inflation and radiation was replaced by a more or less
complex period between these two regimes, period dominated by one or two
fluids defined by equations of state of the form $p=\gamma \varrho $ . All
transitions involved were taken as continuous. Four examples were given,
whose spectra can be seen in figures 1-4.

Some additional comments for the stiff matter case, figure 1, are in order.
A bump in the power-spectrum was found, beginning around $10^{7}$ rad s$^{-1}$, but
not a spike as big as the one reported in ref. 3. This is possibly due to our use of
continuous transitions, automatically avoiding the inevitable
over-production of high-frequency gravitons that take place in the sudden transition
approximation. If we accept, for the scale of inflation, the limit derived
in ref. 13 from WMAP, it is difficult to see how a much larger signal can be
obtained.

Concerning the possibilities of observing these new features, it is
important to stress that it will not be enough to detect the flat part of
the spectrum above $10^{7}$ rad s$^{-1}$, we should also be able to detect the spectrum
below this frequency. Only then we will be certain that a new feature is
indeed present. We shall have to wait for future generations of gravitational-wave detectors, in order to be able to see the main features described in this paper, which take place at frequencies above $10^{7}$ rad s$^{-1}$ (see ref. 14, particularly page 332). The same applies to the other examples we have studied. We agree with the final comment in ref. 3 that the GHz region should be carefully exploited.

\bigskip

\underline{Acknowledgements}

The author wishes to thank Lu\'{i}s E. Mendes for his comments and many
discussions. I am very grateful to Paulo M. S\'{a} for having discovered an error in the programme used to calculate the graphs presented in the previous version of this paper.

\bigskip

\bigskip

\includegraphics{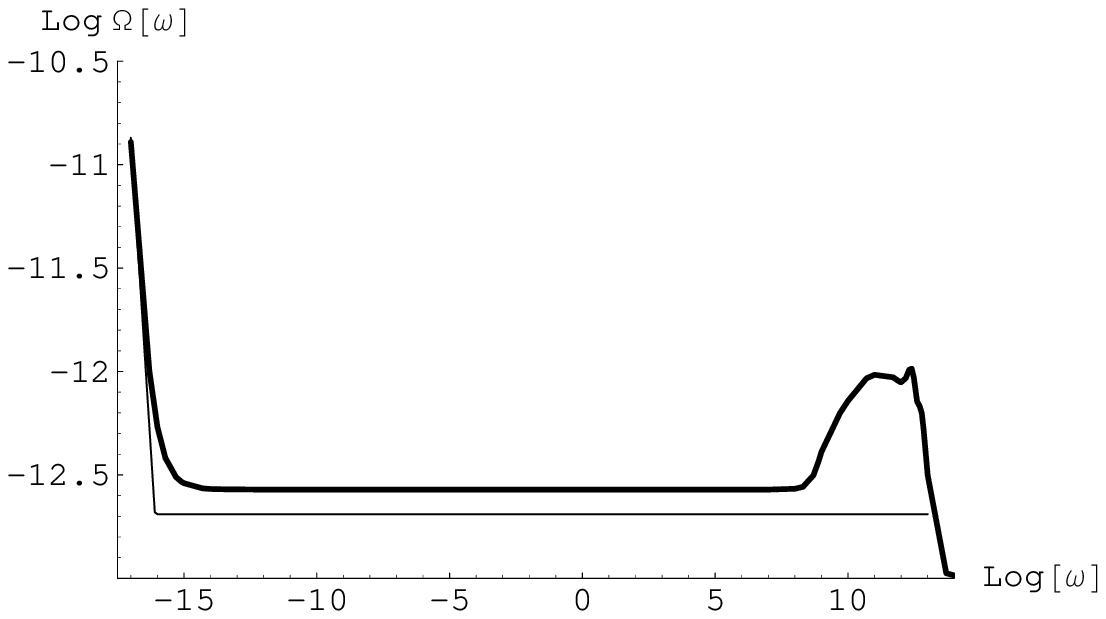}

Fig. 1 - The relative logarithmic energy-spectrum $\Omega$$(\omega)$ corresponding to
example 1. We have inflation followed by a period dominated by a fluid described by a stiff matter
equation of state. The thin curve shows the result for the sudden transition approximation,
as defined in the text. In all figures we have the logarithms in the basis 10.

\bigskip

\bigskip

\includegraphics{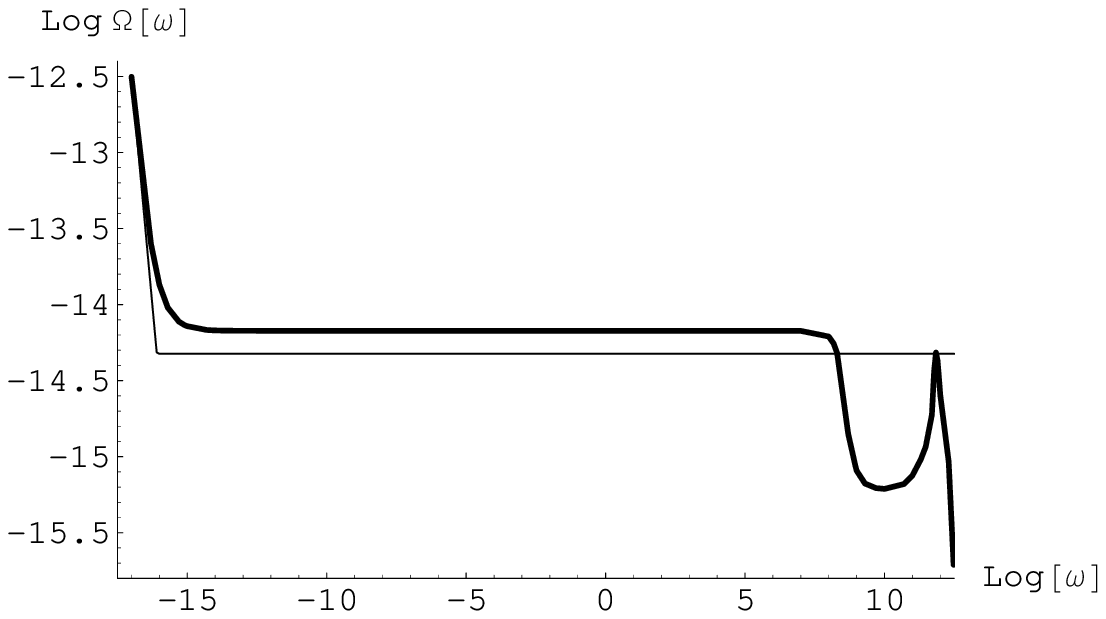}

Fig. 2 - The case described in example 2, where inflation is followed by a period dominated
by a fluid with the dust equation of state $p=0$. The thin line again refers to the
sudden transition approximation.

\bigskip

\bigskip

\includegraphics{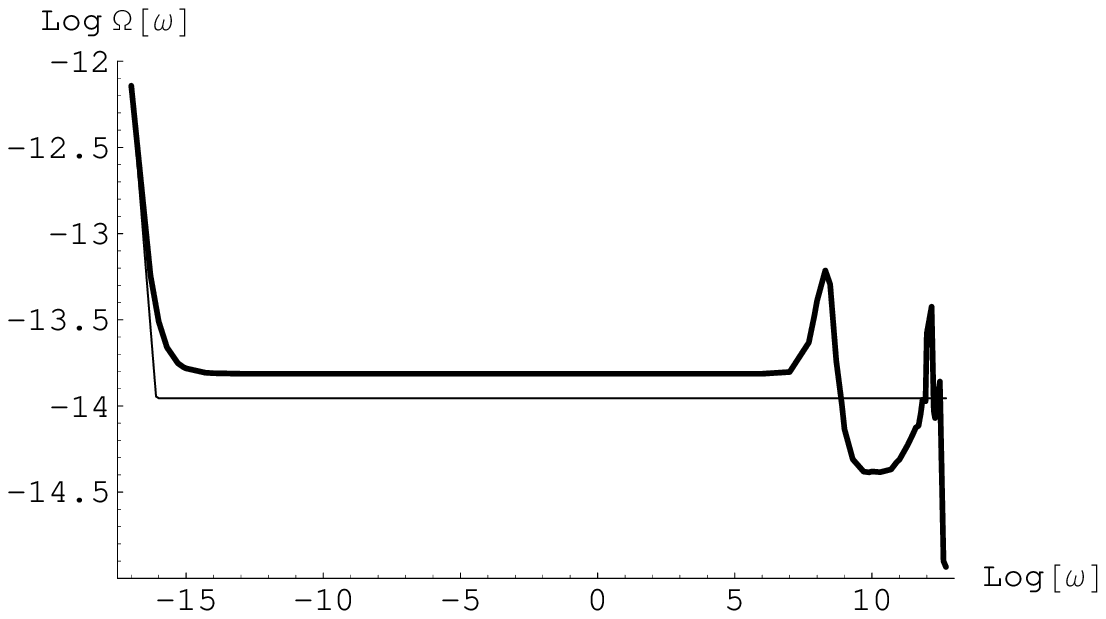}

Fig. 3 - $\Omega$ for the case of example 3, where we have inflation followed by a period
dominated by dust $(p=0)$, in turn followed by a fluid described by a stiff matter equation
of state. The sudden transition results are represented by the thin line.

\bigskip

\bigskip

\includegraphics{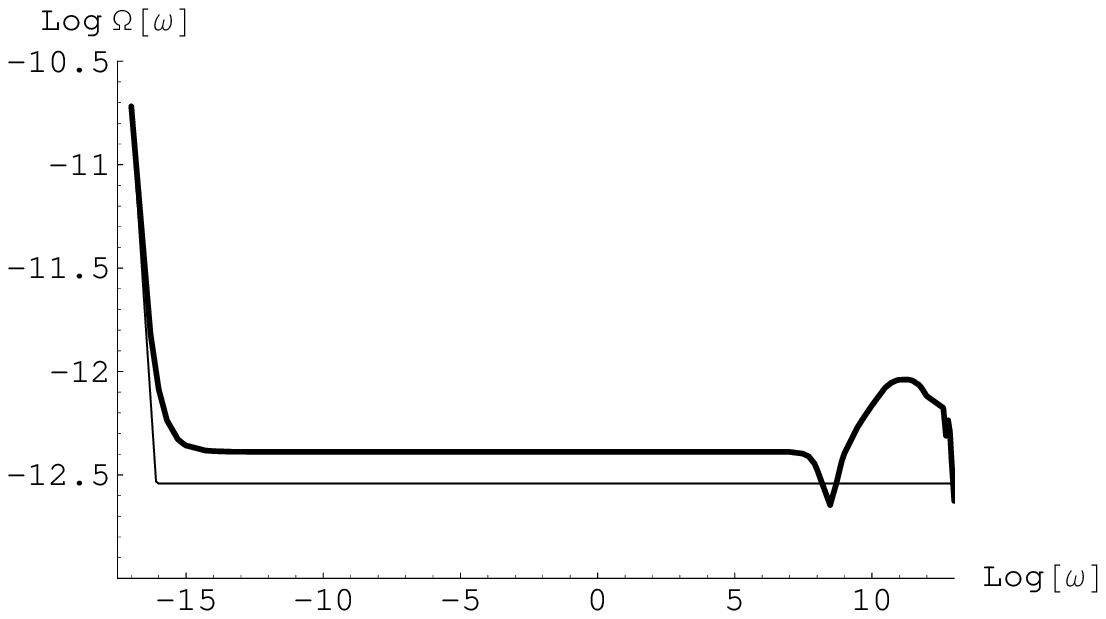}

Fig. 4 - $\Omega$ for the case of example 4, with inflation followed by stiff matter, in turn
followed by dust. Thin line for the sudden transition approximation.

\end{document}